\documentclass{PoS}
\usepackage{amsmath}
\usepackage{mathtools}
\RequirePackage{xspace}
\def\Y#1S{\ensuremath{\Upsilon{(#1S)}}\xspace}
\def\bbbar  {\ensuremath{b\overline b}\xspace}
\def\xbb{\ensuremath{X_{b\overline{b}}}(9975)\xspace}
\def\invfb{\ensuremath{\mbox{\,fb}^{-1}}\xspace}
\newcommand{\gev}{\ensuremath{\mathrm{\,Ge\kern -0.1em V}}\xspace}
\newcommand{\gevcc}{\ensuremath{{\mathrm{\,Ge\kern -0.1em V\!/}c^2}}\xspace}
\newcommand{\mev}{\ensuremath{\mathrm{\,Me\kern -0.1em V}}\xspace}
\newcommand{\mevcc}{\ensuremath{{\mathrm{\,Me\kern -0.1em V\!/}c^2}}\xspace}
\title{Study of Transitions and Decays of Bottomonia at Belle}

\ShortTitle{Study of Transitions and Decays of Bottomonia at Belle}

\author{\speaker{S. Sandilya}\thanks{On behalf of the Belle Collaboration}\\
        University of Cincinnati, Cincinnati, Ohio 45221\\
        E-mail: \email{saurabhsandilya@gmail.com}}

\abstract{We review the studies of transitions and decays of bottomonium 
  states from the $e^{+}e^{-}$ data recorded by the Belle detector at various 
  $\Upsilon$-resonances. 
  We also report a recent study of the branching fractions of $\chi_{bJ}(1P)$ 
  to 41 hadronic final states in the $\Y2S$ data sample. 
  In the same study, a $90\%$ confidence-level upper limit is set for the 
  first time on the width of $\chi_{b0}(1P)$ at $\Gamma_{\rm total}< 2.4\mev$.}

\FullConference{XIII International Conference on Heavy Quarks and Leptons\\
		22-27 May, 2016\\
		Blacksburg, Virginia, USA}

\begin{document}

\section{Introduction}
\label{sec:intro}
Bottomonium, a bound system of a bottom ($b$) quark and its antiquark ($\bar{b}$), offers a unique laboratory to study the strong interaction; since the $b$ quark is heavier than other quarks ($q = u,d,s,c$), the system can be described by non-relativistic quantum mechanics and effective theories~\cite{Brambilla:2012cs}.

The typical radius of the bottomonium from the potential model is of the order of $10^{-1}fm$, and with such a small radius the idea of multipole expansion (widely used in electrodynamics) can be applied to the soft gluon emissions in hadronic transitions between different bottomonium states~\cite{multipole}. 
In the case of radiative transtions, the wavelength of the photon is larger (atmost comparable) to the size of radiating system hence dipole transitions dominate. 

This proceeding presents a review of studies of radiative and hadronic transitions between various bottomonium states as well as their decays to several hadronic final states, based on the data recorded with the Belle detector located at the KEKB asymmetric-energy $e^{+}e^{-}$ collider. 
The $\Upsilon(nS)$ states have $J^{PC}=1^{--}$, hence it can be directly produced in the $e^{+}e^{-}$ collisions when the total center-of-mass (CM) energy is close to the resonance mass. 
And, other bottomonium states are mostly originate from the $\Upsilon(nS)$ decays.

\section{Dipion transitions from  {\boldmath \Y5S}}
\begin{figure}[htp]
  \centering
  \includegraphics[scale=0.65]{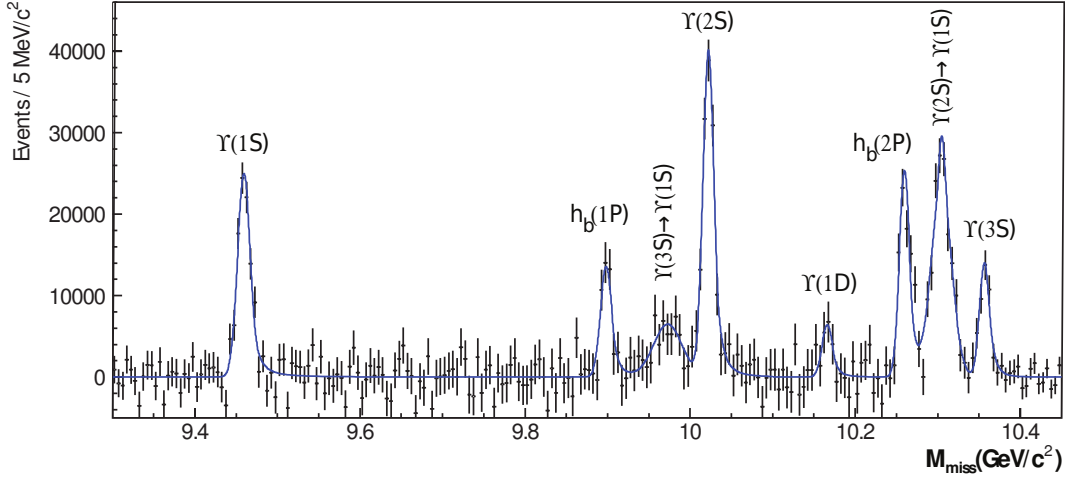}
  \caption[Missing mass spectrum against the dipion system in data recorded with Belle at the \Y5S resonance.]{Missing mass spectrum against the dipion system in the data recorded with Belle at the \Y5S resonance. The plot is taken from Ref.~\cite{belle:hb}.}
  \label{fig:hb}
\end{figure}

The Belle Collaboration observed two orders of magnitude larger partial width of the dipion transition $\Y5S \to \pi^{+}\pi^{-} \Upsilon(mS)$, where $m = 1,~2$ and 3 than the corresponding partial widths from \Y4S, \Y3S, or \Y2S~\cite{chen}. The dipion transition from \Y5S via intermediate $B^{(*)}{\bar B^{(*)}}$ re-scattering can explain this large branching fraction~\cite{meng}. 
Another interesting result in this transition from \Y5S was the first observation of $h_{b}(1,2P)$~\cite{belle:hb}. In that analysis,
Belle observed that cross-sections of the processes  
$\Y5S \to \pi^{+}\pi^{-} \Y2S$ and $\Y5S \to \pi^{+}\pi^{-} h_{b}(1,2P)$
are of comparable magnitude, as shown in Figure~\ref{fig:hb}. Such a large 
rate was unexpected because production of $h_{b}(1,2P)$ requires a $b$-quark 
spin flip, while that of \Y2S does not. 
This indicates that the production of $h_{b}(1,2P)$ at \Y5S must occur via a 
process that mitigates the expected suppression related to heavy quark spin, 
which subsequently led to the observation of two charged bottomoniumlike 
resonances~\cite{belle:zb}.

\section{Radiative transitions from {\boldmath $h_{b}(1,2P)$}}
\begin{figure}[htp]
  \centering
  \includegraphics[scale=0.65]{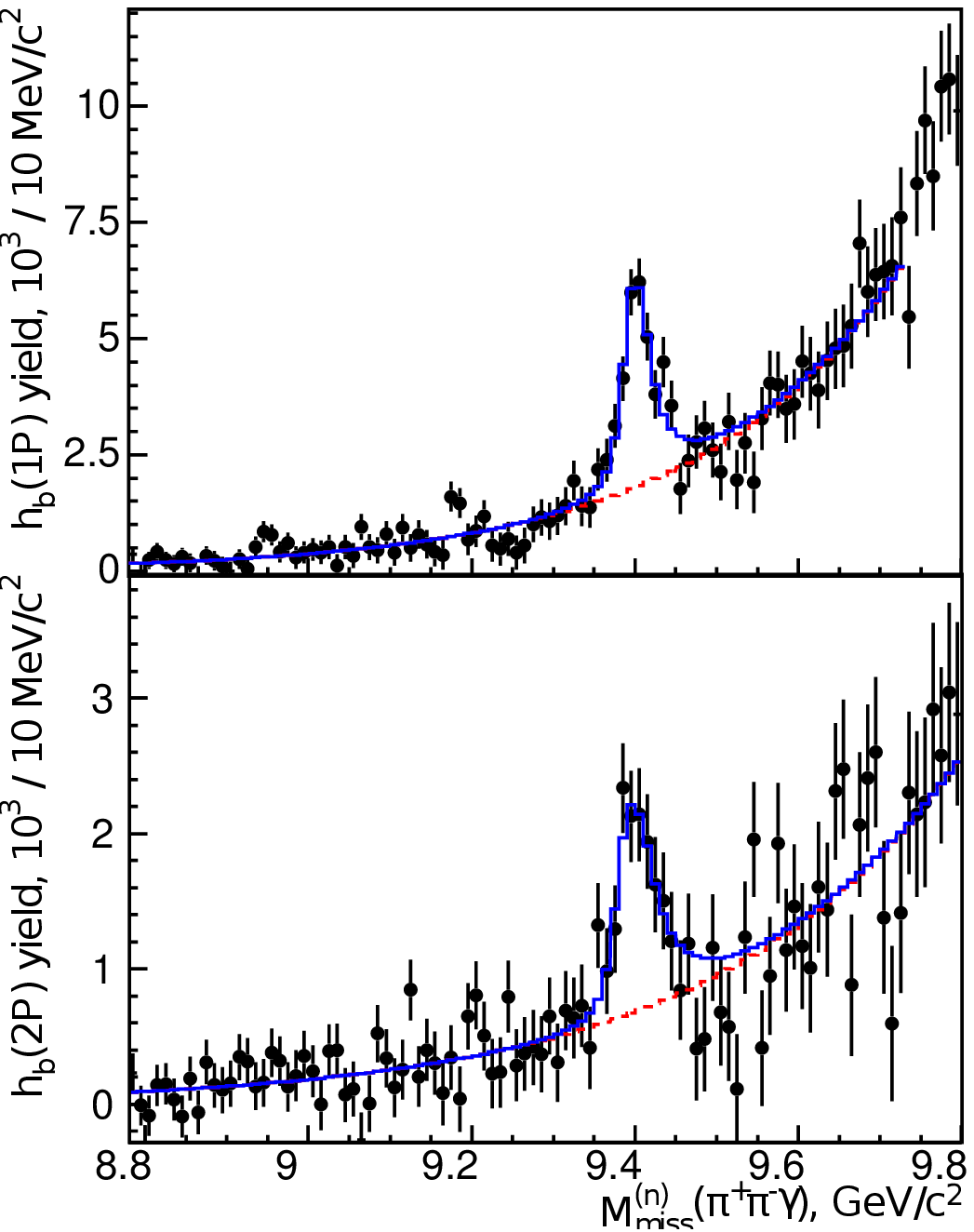}
  \includegraphics[scale=0.65]{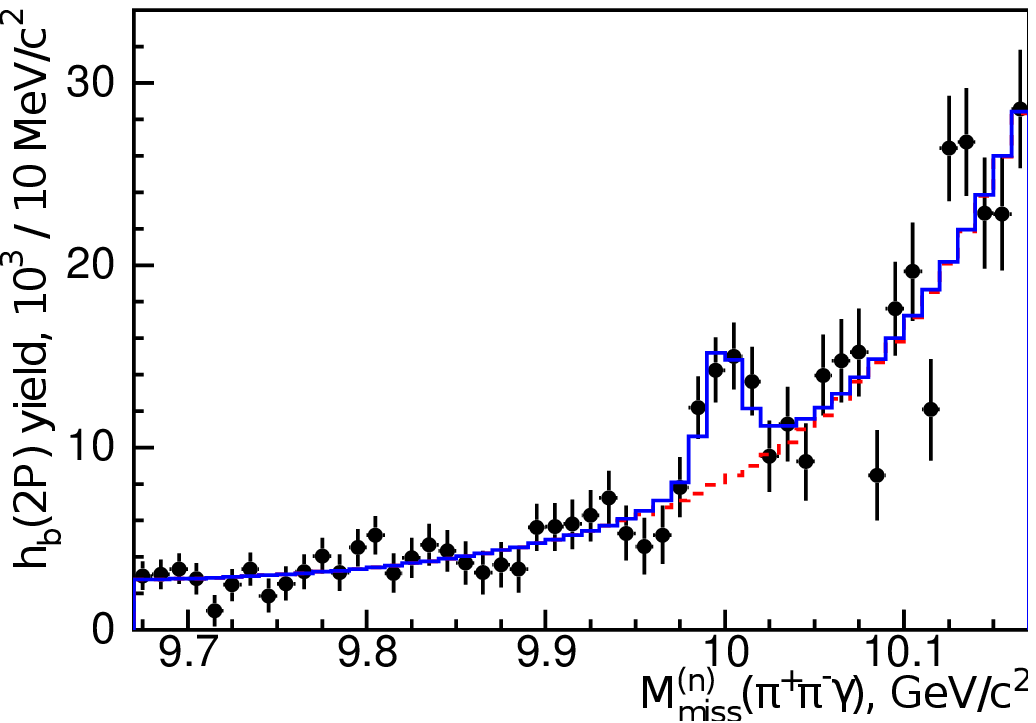}
  \caption{$h_{b}(1P)$ yield {\it vs.} $M_{miss}^{(1)}(\pi^{+}\pi^{-}\gamma)$ (top-left), and $h_{b}(2P)$ yield {\it vs.} $M_{miss}^{(2)}(\pi^{+}\pi^{-}\gamma)$ distribution in the $\eta_{b}(1S)$ (bottom-left) and $\eta_{b}(2S)$ (bottom-right) regions. Points with error bars are the data, blue solid histograms are the result from a fit, and the red dashed histograms are the background component of the fit. The plot is taken from Ref.~\cite{belle:e2s}.}
  \label{fig:e12smiz}
\end{figure}

A 133.4 \invfb of data sample collected near the \Y5S resonance is used to study  the process $e^{+}e^{-}\to\Y5S\to h_{b}(nP)\pi^{+}\pi^{-},{\bf \it h_{b}(nP)\to\eta_{b}(mS)\gamma}$ for $n\geq m = 1{\rm~and~}2$, in which the $\eta_{b}(mS)$ states are not exclusively reconstructed~\cite{belle:e2s}. 
In this analysis, Belle reported the first evidence for the $\eta_{b}(2S)$ as well as the most precise mass measurement of $\eta_{b}(1S)$ and its width for the first time. 
To identify signal events, only a photon and $\pi^{+}\pi^{-}$ pair are reconstructed and a missing mass against \Y5S is defined as 
$M_{miss}(X)=\sqrt{(E_{\rm CM}-E_{X}^{\star})^{2}-p_{X}^{\star 2}}$, where $E_{\rm CM}$ is the CM energy, and $E^*_X$ and $p^*_X$ are the energy and momentum of the recoiling system $X$ measured in the CM frame. 
The missing mass against the dipion system, $M_{miss}(\pi^{+}\pi^{-})$, is used in selecting $h_{b}(nP)$
signal events, while the $\eta_{b}(mS)$ signal is identified with the variable,
$M_{miss}^{(n)}(\pi^{+}\pi^{-}\gamma) = M_{miss}(\pi^{+}\pi^{-}\gamma)-M_{miss}(\pi^{+}\pi^{-})+m_{h_{b}(nP)}$. 
The $h_{b}(nP)$ yield {\it vs.} $M_{miss}^{(n)}(\pi^{+}\pi^{-}\gamma)$ distribution is shown in Figure~\ref{fig:e12smiz} (left plots), where the peak near 9.4\gevcc is identified as the $\eta_{b}(1S)$. The mass of the $\eta_{b}(1S)$ state is measured as $9402.4\pm 1.5\pm 1.8$ \mevcc, which corresponds to the hyperfine splitting of [$\Delta M_{HF}(1S)$] $57.9\pm 2.3$\mevcc. The $\eta_{b}(1S)$ width is also measured in the same analysis and found to be $10.8\substack{+4.0 \\ -3.7}\substack{+4.5 \\ -2.0}$ \mev.

The $h_{b}(2P)$ yield {\it vs.} $M_{miss}^{(2)}(\pi^{+}\pi^{-}\gamma)$ distribution is shown in Figure~\ref{fig:e12smiz} (right), where the peak near 10.0\gevcc is identified as the $\eta_{b}(2S)$. The significance of the $\eta_{b}(2S)$ signal is $4.2\sigma$ and its mass is $9999.0\pm3.5^{+2.8}_{-1.9}$\mevcc, which corresponds to a hyperfine splitting for $2S$, $\Delta M_{HF} = 24.3^{+4.0}_{-4.5}$\mevcc. 

\section{Radiative transitions from \Y2S}
\label{sec:rady2s}

\begin{figure}[htp]
\includegraphics[scale=0.8]{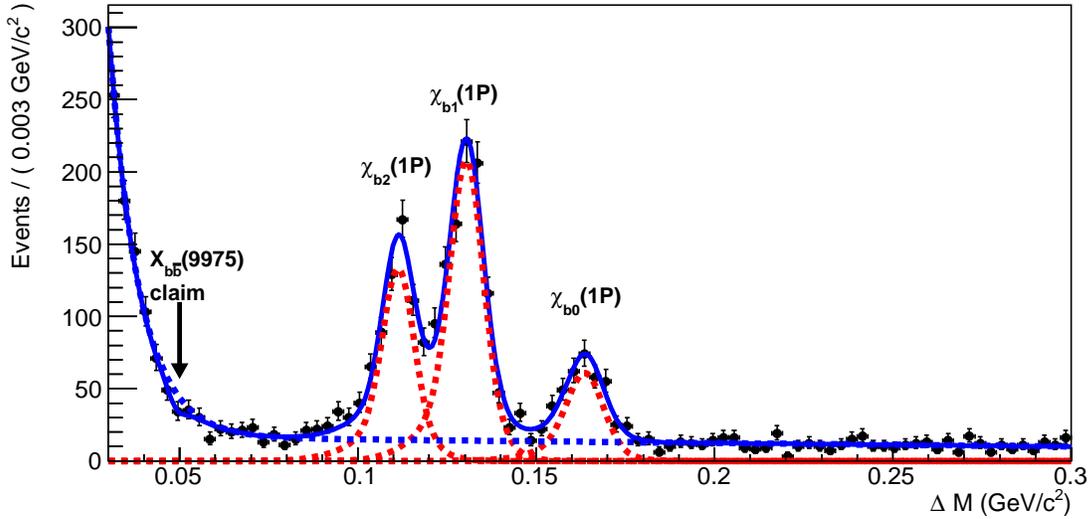}
\caption{$\Delta M$ distributions for \Y2S data events. Points with error bars are the data, the blue solid curve is the result of the fit for the signal-plus-background hypothesis, and the blue dashed curve is the background component. The plot is taken from Ref.~\cite{belle:xbb}.}
\label{fig:e2s}
\end{figure}

The data for $158 \times 10^{6}$ \Y2S decays recorded with the Belle detector have been used to study the radiative population of states reconstructed from 26 different exclusive hadronic final states~\cite{belle:xbb}. 
This analysis was done with the motivation to search for a bottomonium state near 9975 \mevcc, \xbb, claimed to have been observed in the radiative \Y2S decays recorded by the CLEO detector~\cite{Dobbs}. 
We found no evidence for an \xbb signal; The reconstructed invariant mass of the \bbbar system is presented in terms of $\Delta M\equiv M[(\bbbar)\gamma]-M(\bbbar)$, and result of the fit to the data is shown in Figure~\ref{fig:e2s}. 
A 90\% CL upper limit is determined on the product branching fraction $ {\cal B}[ \Y2S \to \xbb \gamma ] \times \sum_i { \cal B }[ \xbb \to h_{i} ] <  4.9 \times 10^{-6} $, which is an order of magnitude smaller than that reported in Ref.~\cite{Dobbs}. 

The disconfirmation of the \xbb state was extremely important as it was claim to be the $\eta_{b}(2S)$ state was in disagreement with theory predictions and the Belle result~\cite{belle:e2s}. 
We have also searched for the $\eta_{b}(1S)$ state and set an upper limit ${\cal B }[ \Y2S \to \eta_{b}(1S) \gamma ] \times \sum_i {\cal B }[\eta_{b}(1S) \to h_i ]< 3.7 \times 10^{-6}$ at 90\% CL. 
A large signal yield is observed for $\chi_{bJ}(1P)$ states from the sum of 26 exclusive hadronic final states, in the same analysis. 
The yields for $\chi_{bJ}(1P)(J=0,1,2)$ were $299 \pm 22$, $946 \pm 36$ and $582 \pm 31$, respectively. 

\section{\bf $\chi_{bJ}(1P)$ decays to light hadrons}
The analysis mentioned in the Section~\ref{sec:rady2s}, motivated a study of the product branching fractions 
${\cal B}[\Y2S\to\gamma\chi_{bJ}(1P)]\times{\cal B}[\chi_{bJ}(1P)\to h_{i}]$, 
where $h_{i}$ is a specific hadronic mode; such decays of $\chi_{bJ}(1P)$ mesons
give us insight into how initial quarks and gluons hadronize~\cite{patrignani1}.
We performed this study using the same $\Y2S$ datasets with the $\chi_{bJ}(1P)$ being copiously produced from the \Y2S via electric dipole radiative transitions. 
The $\chi_{bJ}(1P)$ states can decay to many hadronic final states. For decays into all-charged final states, we focus on the same $26$ modes as in Ref.~\cite{belle:xbb}. 
In addition, modes with one $\pi^{0}$ and two $\pi^{0}$'s are added to the mentioned final states excluding $2(\pi^+\pi^-)\pi^0$, $3(\pi^+\pi^-)\pi^0$, $4(\pi^+\pi^-)\pi^0$ and $5(\pi^+\pi^-)\pi^0$ as those are forbidden by G-parity conservation). 
In total, 74 light hadronic decay modes of the $\chi_{bJ}(1P)$ are reconstructed. 
A detailed analysis procedure of this study is mentioned in Ref.~\cite{belle:chibj}. 

\begin{figure}[htb]
  \centering
  \includegraphics[width=0.8\textwidth]{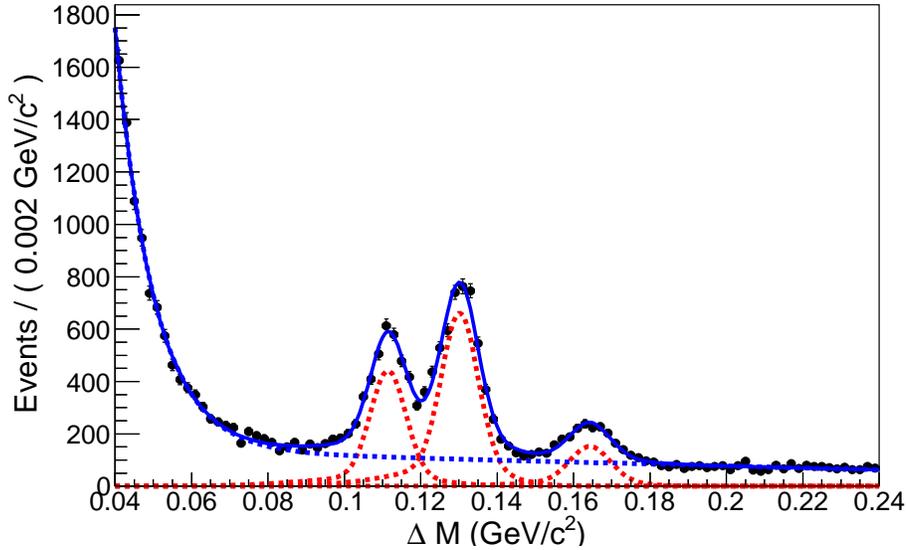}
  \caption{$\Delta M$ distribution in the $\Y2S$ data. The three $\chi_{bJ}(1P)$ ($J=0,1,$ and 2, respectively from right to left) components are indicated by the red dashed curves. The plot is taken from Ref.~\cite{belle:chibj} }
  \label{fig:fitdt}
\end{figure}

The result of the likelihood fit to the $\Delta M$ ($\Delta M\equiv M[\chi_{bJ}(1P)\gamma]-M[\chi_{bJ}(1P)]$) distribution for the sum of $74$ modes in data is shown in Figure~\ref{fig:fitdt}. The masses of the $\chi_{bJ}(1P)(J=0,1,2)$ states obtained are in good agreement with their world average values~\cite{PDG}. 

In total, 41 modes have above $5$ standard deviation ($\sigma$) significance in at least one of the $\chi_{bJ}(1P) (J=0,1,2)$ signals. 
Our branching fraction results are consistent with, and more precise than, reported in a similar analysis performed in data collected by the CLEO III detector~\cite{cleo:chibj}. 
Furthermore, a $\chi_{bJ}(1P)$ signal for $J=0,1$, and $2$ has been observed for the first time in 9, 27, and 16 modes, respectively. 
And, we also found first evidence of a $\chi_{bJ}(1P)$ signal in 18, 1, and 14 modes for $J=0,1$, and $2$, respectively. 

The large signal yield obtained in our branching fraction studies of the 
$\chi_{bJ}(1P)$ triplet motivates a width measurement of the $\chi_{b0}(1P)$.
The $\chi_{b0}(1P)$ width is found to be 1.3 $\pm$ 0.9\,\mev.
In the absence of a statistically significant result, we derive an upper limit 
on the width of the $\chi_{b0}(1P)$ $< 2.4 \mev$ at $90\%$ CL.

\section{\bf $\eta$ transition from \Y4S}
As mentioned in Section~\ref{sec:intro}, hadronic transitions between the lower mass quarkonium levels can be described using the QCD mulitpole expansion. 
The $\pi\pi$ and $\eta$ transtions between the vector states proceed via emission of $E1E1$ and $E1M2$ gluons, respectively. Therefore, the $\eta$ transitions are highly suppressed as they require a spin flip of the heavy quark~\cite{voloshin}. The ratio of branching fractions:
\[
\mathcal{R}^{\eta S}_{\pi\pi S}(n,m)=
\frac{\mathcal{B}[\Upsilon(nS)\to\eta\Upsilon(mS)]}
{\mathcal{B}[\Upsilon(nS)\to\pi^+\pi^-\Upsilon(mS)]}
\]
is measured to be small for low-lying states:
$\mathcal{R}^{\eta S}_{\pi\pi S}(2,1)=
(1.64\pm0.23)\times10^{-3}$~\cite{He:2008xk, BABAR:2011ab, Tamponi:2012rw} and
$\mathcal{R}^{\eta S}_{\pi\pi S}(3,1)<2.3\times10^{-3}$~\cite{BABAR:2011ab}.

Above the $B\bar{B}$ threshold, BaBar unexpectedly observed the transition
$\Upsilon(4S)\to\eta\Upsilon(1S)$ with a large branching fraction of
$(1.96\pm0.28)\times10^{-4}$, corresponding  to
$\mathcal{R}^{\eta S}_{\pi\pi S}(4,1)=2.41\pm0.42$~\cite{Aubert:2008az}.
This apparent violation of the heavy quark spin-symmetry was explained
by the contribution of $B$ meson loops or, equivalently, by the
presence of a four-quark $B\bar{B}$ component inside the $\Upsilon(4S)$
wave function~\cite{Meng:2008,Voloshin:2011}.
At the $\Upsilon(5S)$ energy, the anomaly is even more striking. 
The spin-flipping processes $\Upsilon(5S)\to \pi\pi h_b(1P, 2P)$ have been found not to be suppressed with respect to  the spin-symmetry  preserving reactions $\Upsilon(5S)\to \pi\pi \Upsilon(1S, 2S)$ \cite{belle:hb}, and all the $\pi\pi$ transitions showed the presence of new resonant structures \cite{Bondar:2012,Krokovny:2013} that cannot be explained as conventional bottomonium states. 

\begin{figure}[htb]
  \centering
  \includegraphics[scale=0.65]{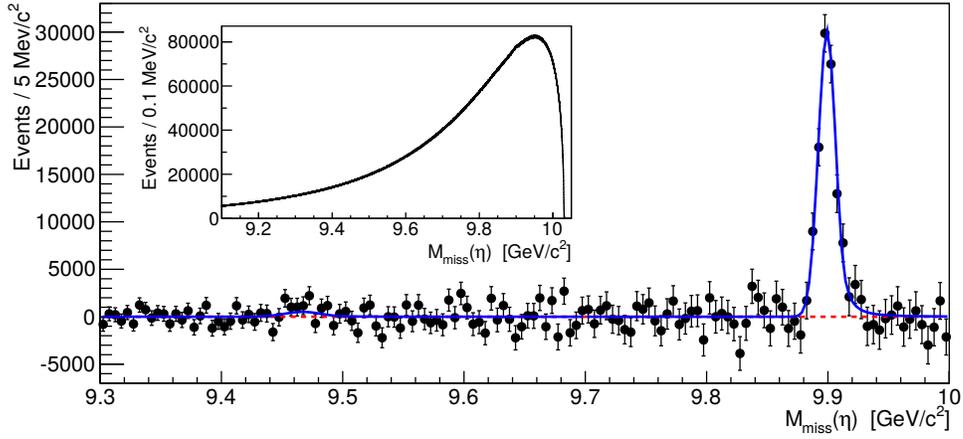}
  \caption{$M_{\mathrm{miss}}(\eta)$ distribution after the background  subtraction. 
    The blue solid curve shows the fit with the signal PDFs, while the red-dashed line represents the background only hypothesis. 
    The inset shows the $M_{\mathrm{miss}}(\eta)$ distribution before the background subtraction. 
    The plot is taken from Ref.~\cite{Tamponi:2015}}
   \label{fig:eta_hb}
\end{figure}

Further insight into the mechanism of the hadronic transitions above the threshold can be gained by searching for the $E1M1$ transition $\Upsilon(4S)\to\eta h_b(1P)$, which is  predicted to have a branching fraction of the order of $10^{-3}$~\cite{Guo:2010ca}. 
The missing mass spectrum of $\eta$ meson, $M_{\rm miss}(\eta)=\sqrt{(p_{e^+e^-}-p_{\eta})^2}$, where $p_{e^+e^-}$ and $p_{\eta}$ are, respectively, the four-momenta of the colliding $e^+e^-$  pair and the $\eta$ meson, is investigated in the $772 \times 10^{6}$ \Y4S decays collected by Belle ~\cite{Tamponi:2015}. 
Figure~\ref{fig:eta_hb} shows the background-subtracted $M_{\mathrm{miss}}(\eta)$ distribution, where the transition $\Y4S \to \eta h_b(1P)$ is observed for the first time with a statistical significance of $11 \sigma$. 
The branching fraction of the transition, ${\cal B}[\Upsilon(4S) \to \eta h_b(1P)] = (2.18 \pm 0.11 \pm 0.18 ) \times 10^{-3}$, is in agreement with the theoretical prediction \cite{Guo:2010ca}. 

In the same analysis,  the $\eta_b(1S)$ is also studied by reconstructing the transitions $\Upsilon(4S) \to \eta h_b(1P) \to \eta \gamma \eta_b(1S)$. The  $\eta_{b}(1S)$ signal is observed with a statistical significance of 9$\sigma$, which allowed to measure its mass  $9400.7 \pm 1.7 \pm 1.6 \mevcc $ and width, $\Gamma_{\eta_b(1S)} = (8 ^{+6}_{-5} \pm 5) \mev$.

\section{{\boldmath $\Upsilon(1,2S)$} decays to {\boldmath $\Lambda\bar{\Lambda}$}}
\label{sec:lambda}
The inclusive production rate of $\Lambda$ has been found to be three times larger in the \Y1S annihilation with respect to the nonresonant $e^{+}e^{-}\to q \bar{q}$ events at the same CM energy~\cite{Briere:2007}. 
An analysis is performed in the \Y1S and \Y2S datasets collected with the Belle detector to study the exclusive production of hyperons or, specifically $\Lambda\bar{\Lambda}$ pair~\cite{Tamponi:20XX}. 
The \Y1S and \Y2S are reconstructed in 48 final states comprising a $\Lambda\bar{\Lambda}$, up to three $p\bar{p}$, $K^{+}K^{-}$ or $\pi^{+}\pi^{-}$ pairs, and one $\pi^{0}$.

\begin{figure}[htb]
  \centering
  \includegraphics[scale=0.45]{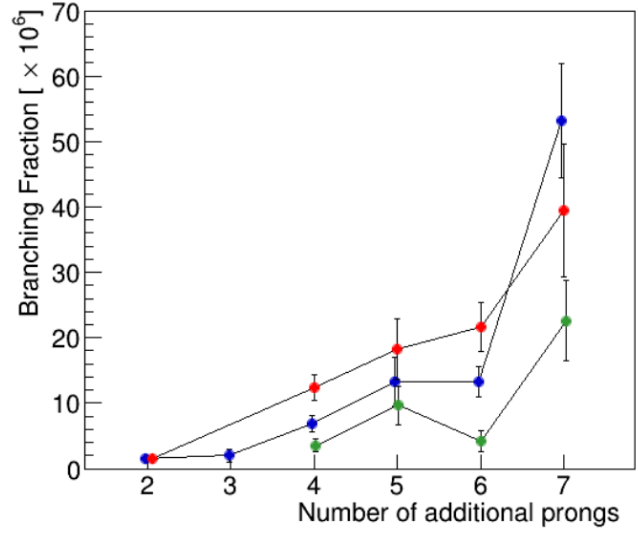}
  \caption{Branching fraction of $\Y1S\to\Lambda\bar{\Lambda}+n\pi$ (red dots), $\Y2S\to\Lambda\bar{\Lambda}+n\pi$ (green dots) and $\Y1S\to\Lambda\bar{\Lambda}K^{+}K^{-} + (n-2)\pi$ (blue dots) as a function of total number of mesons (n) produced in association with the $\Lambda\bar{\Lambda}$ pair. 
     The plot is taken from Ref.~\cite{Tamponi:20XX}}
   \label{fig:lambda}
\end{figure}

In total, we observe 16 \Y1S and 7 \Y2S annihilation modes with a $\Lambda\bar{\Lambda}$ pair. Furthermore, we concluded that the $\Lambda\bar{\Lambda}$ production occurs mostly in high multiplicity environment. Also, shown in Figure~\ref{fig:lambda}, the branching fraction scales with the increasing number of additional particles. 

\section{Summary}
It was an exciting decade for heavy quarkonium physics specially at the Belle experiment, with discoveries of a number of conventional and non-conventional bottomonium states. In this proceeding, only few recent interesting results of transitions and decays of bottomonium have been summarized. We also reported a recent study of the branching fractions of $\chi_{bJ}(1P)$ to several hadrons in the $\Y2S$ data sample. In the same study a $90\%$ CL upper limit is set on the width of $\chi_{b0}(1P)$ at $\Gamma_{\rm total}< 2.4\mev$.

\end{document}